\documentstyle[prd,aps,eqsecnum]{revtex}
\begin{document}
\title{Signature of Chaos in Gravitational Waves from a Spinning Particle}
\author{{\sc Shingo} SUZUKI\thanks{electronic
mail:ssuzuki@th.phys.titech.ac.jp}}
\address{Department of Physics, Tokyo Institute of Technology,
Meguro-ku, Tokyo 152, Japan}
\author{{\sc Kei-ichi}
 MAEDA$^{1,2}$\thanks{electronic mail:maeda@mse.waseda.ac.jp}}
\address{$^1$ Department of Physics, Waseda University,
Shinjuku, Tokyo 169-8555, Japan\\[.3em]
$^2$ Advanced Research Institute for Science and Engineering,\\
Waseda University, Shinjuku, Tokyo 169-8555, Japan}
\date{\today}
\maketitle
%
\baselineskip 18pt
\begin{abstract}
A spinning test particle around a
Schwarzschild
black hole shows a chaotic behavior, if its spin is larger than a critical
value. We discuss whether or not some peculiar signature of chaos appears in
the gravitational waves emitted from such a system. Calculating the
 emitted gravitational waves
 by use of the
quadrupole formula, we find that the energy emission rate of gravitational
waves for a chaotic orbit is  about 10 times larger than that for a circular
orbit, but the same enhancement is also obtained by a  regular ``elliptic"
orbit. A chaotic motion is not always  enhance the
energy emission rate maximally. As for the energy spectra of the
gravitational waves, we find some characteristic feature for
a chaotic orbit. It may tell us how to find  out a  chaotic
behavior of the system.
Such a peculiar behavior, if it will be found, may also provide us some
additional informations to determine  parameters of a system such as
a spin.
\end{abstract}
%
\baselineskip 10mm
\section{Introduction}
\label{sec1}\setcounter{equation}{0}
Since the study by Poincar\'e about three-body problem, many works about chaos have
been done. 
In our universe, there exist lots of chaotic systems, some of them are actually observed. 
For example, the motion of Hyperion is probably to
be chaotic\cite{Klavetter}. 
The gaps in distribution of the asteroids between Mars and Jupiter may
be formed in consequence of a chaotic mechanism\cite{Moser}. 
In addition, some signals from the Universe contain the signature of
chaos\cite{Lehto}--\cite{Kollath}. 
In these literature, the authors concluded that the irregular variation of the light
curve of optical light or X-ray from some astrophysical objects such as a quasar
or a black hole are governed by chaotic dynamics, and they have claimed that these chaotic signals help us to see the internal
stellar structure of the stars. 

Gravitational waves bring us various information about the
Universe or astrophysical objects. 
It would be detected directly on the Earth hopefully at the beginning of the next
century by laser interferometric detectors, such as the U.S. Laser
Interferometric Gravitational Wave Observatory(LIGO)\cite{LIGO}. 
The most promising sources of the gravitational waves for these detectors are
the binary systems consisting of neutron stars and/or black holes. 
If we detect a signal of gravitational waves emitted from those systems and
compare its wave form with that obtained theoretically (templates), we may be able
not only to determine a variety of astrophysical parameters of the sources, for
example their orbital information, masses, spins and so on\cite{Droz}, but also
to obtain more information about fundamental physics, for example the equation
of state at high density\cite{Cutler}. 
This is the so-called gravitational wave astronomy, which would be one of the most important subjects in the next century. 

To extract a correct information of the source from the detected signals,
it is important to make the exact templates of the gravitational waves.
To do that, we need to know the exact motion of the system.
Because the gravitational field around the binary system is very strong,
the motion of the system should be calculated not in the Newtonian theory of
gravity but in Einstein's theory of gravity.
Hence numerical relativity is one of the most promising treatments to make correct
templates, but it is still at a developing stage.
Instead, some approximation methods, such as the black hole perturbation method
or the post-Newtonian approximation, are used to make the
templates\cite{Damour}.
From these works, we can see that some general relativistic effects make the motion
complicated.  Due to such effects, the motion of the system may show a chaotic behavior, which do not exist in the Newtonian theory
of gravity.

 In this paper, we pay a special attention to the relativistic
spin interaction, because we know that the spin--orbit and spin--spin
interaction plays a very important role in the gravitating system. 
For example, in \cite{Apostolatos}, the motion of the binary system
consisting of spinning neutron stars is analyzed
using the post-Newtonian approximation. 
They showed that the orbital plane of the system behaves in a strange way
and the gravitational waves are modified through it. 

The gravitational waves emitted from a spinning test particle have been calculated by several authors\cite{Mino},\cite{Saijo}. 
In \cite{Mino}, the gravitational waves produced by a spinning test particle falling into a Kerr black hole along the symmetric axis of the spacetime or moving circularly around it ware discussed and the energy emission rate from those systems was calculated. 
The gravitational waves emitted from a spinning test particle plunging into a Kerr black hole from infinity were calculated in
\cite{Saijo}. 

The motion of a spinning test particle in a relativistic spacetime can be complicated in general. 
In our previous paper\cite{ShingoI}(hereafter, referred to as Paper I),
the general motion of a spinning test particle around a Schwarzschild black
hole was analyzed.
We found that the spin--orbit interaction makes some orbits chaotic.
In \cite{ShingoII}(hereafter, referred to as Paper II), the
spin effect of the particle on the stability of a circular orbit was
investigated, and a new type of instability of the circular orbit was found.
We showed that if this instability occurs in the binary system, the
innermost stable circular orbit (ISCO), which is one of the most important
keys in gravitational wave observation, would change
and the orbital plane of the system will precess at the final moment of the
inspiral phase.

Thus, taking into account the relativistic effects on the particle
motion, the motion of the particle can be changed drastically.
Then the emitted gravitational waves in the chaotic system can be different from those in the regular system. 
If the binary system behaves in a chaotic way and we observe it through
gravitational waves, a new information about the source, for example the
lower bound of the spin amplitude, could be obtained.

Therefore, it is important to investigate the chaos in binary systems
and its effects on the gravitational waves, not only from academic
interest but also from the observational point of view.

This paper is organized as follows.
In Sec. II, the model and the basic equations used in the present analysis are
summarized.
In Sec. III, the gravitational waves from the system are calculated using the
quadrupole formula and the differences between the regular and chaotic case
are discussed.
A summary and some remarks follow in Sec. IV.

Throughout this paper we use units $G=c=1$ and notation including the
signature of the metric follows Misner, Thorne and
Wheeler\cite{Gravitation}.

\section{Basic Equations for a Spinning Test Particle}
\label{sec2}\setcounter{equation}{0}
The equations of motion of a spinning test particle in a
relativistic spacetime were first derived by Papapetrou\cite{Papapetrou}
and then reformulated by Dixon\cite{Dixon}.
Those are a set of equations:
\begin{eqnarray}
\frac{\mbox{d}x^{\mu}}{\mbox{d}\tau}&=&v^{\mu},
\label{eqn:xdot}
\\
\frac{\mbox{D}p^{\mu}}{\mbox{d}\tau}&=&
-\frac{1}{2}R^{\mu}_{~\nu\rho\sigma}v^{\nu}S^{\rho\sigma},
\label{eqn:pdot}
\\
\frac{\mbox{D}S^{\mu\nu}}{\mbox{d}\tau}&=&p^{\mu}v^{\nu}-p^{\nu}v^{\mu},
\label{eqn:sdot}
\end{eqnarray}
where $\tau, v^{\mu}, p^{\mu}$ and $S^{\mu\nu}$ are an affine
parameter of the orbit, the 4-velocity of the particle, the
momentum, and
the spin tensor, respectively.
$\tau$ is chosen as the proper time of the particle in this paper, then $
v^{\mu} v_{\mu} = -1$.
The multipole moments of the particle higher than a spin
dipole are ignored.
It is called the pole-dipole approximation.

Here, using the ``center of mass'' condition, $p_{\mu}S^{\mu\nu}=0$\cite{Dixon}, we obtain the relation between $v^{\mu}$ and
$p^{\mu}$ as
\begin{equation}
v^{\mu}=N\left[u^{\mu}+\frac{1}{2\mu^2\Delta}S^{\mu\nu}u^{\lambda}
R_{\nu\lambda\rho\sigma}
S^{\rho\sigma}\right],
\label{eqn:p-v}
\end{equation}
where
\begin{equation}
\Delta=1+\frac{1}{4\mu^2}R_{\alpha\beta\gamma\delta}
S^{\alpha\beta}S^{\gamma\delta},
\end{equation}
and
\begin{equation}
N=\left[ 1- \frac{1}{4\Delta^2\mu^2} S_{\mu \nu} u_{\lambda}
S_{\rho\sigma}R^{\nu\lambda\rho\sigma}S^{\mu \alpha} u^{\beta}
S^{\gamma\delta}R_{\alpha\beta\gamma\delta}
\right]^{-1/2}
\end{equation}
 is a normalization constant fixed by
$v_{\nu}v^{\nu}=-1$.
 $u^{\nu} \equiv p^{\nu}/\mu$ is a unit vector parallel to the
momentum
$p^{\nu}$, where the mass of the particle $\mu$ is defined by
\begin{equation}
\mu^2=-p_{\nu}p^{\nu}.
\end{equation}

As for a back ground spacetime, we assume a Schwarzschild black hole in this paper.  
The metric is given as
\begin{equation}
\mbox{d}s^2=-f\mbox{d}t^2+f^{-1}\mbox{d}r^2+r^2(\mbox{d}\theta^2+\sin^2\theta \mbox{d}\phi^2),
\end{equation}
\begin{equation}
f=1-\frac{2M}{r},
\end{equation}
where $M$ is the mass of the black hole. 
In this case, we have several conserved quantities, which are 
\begin{eqnarray}
{\cal S}^2&=&\frac{1}{2}S_{\mu\nu}S^{\mu\nu},\label{eqn:spin}\\
\mu^2&=&-p_{\nu}p^{\nu},\label{eqn:mass}\\
E&=&-\xi_{(t)\nu}p^{\nu}+\frac{1}{2}\xi_{(t)\mu;nu}S^{\mu\nu}=-p_t-\frac{M}{r}S^{tr},\label{eqn:energy}\\
J_z&=&\xi_{(\phi)\nu}p^{\nu}-\frac{1}{2}\xi_{(\phi)\mu\nu}S^{\mu\nu}=p_{\phi}-r(S^{\phi
r}-rS^{\theta\phi}\cot\theta)\sin^2\theta,\label{eqn:jz}
\end{eqnarray}
where $\xi_{(t)}^{\mu}$ and $\xi_{(\phi)}^\mu$ are the timelike and axial Killing vector fields of the back ground spacetime, 
respectively. 
${\cal S}$, $\mu$, $E$ and $J_z$ are interpreted as the magnitude of the spin, the mass, the energy and the $z$ component of the
total angular momentum of the particle, respectively.  In addition, because of the spherical symmetry of the back ground spacetime,
the $x$ and $y$ components of the total angular momentum are also conserved.  And without loss of generality, we can set
$(J_x,J_y,J_z)=(0,0,J)$, where $J$ is the magnitude of the total angular momentum. 

The motion of a spinning test particle around a Schwarzschild black hole and its
chaotic behavior were studied in Paper I. 
Here we summarized our results in below.

To analyze the motion of a test particle, it is convenient to introduce
the ``effective potential'' of the particle. 
For a spinning test particle in Schwarzschild spacetime, it is defined as
\begin{equation}
V_{(\pm)}(r,\theta;J_z,{\cal S})=\mu
f^{\frac{1}{2}}(1+\Lambda_{(\mp)}^2)^{\frac{1}{2}}+\frac{M}{r}\Lambda_{(\mp)
}\left({\cal
S}^2-\frac{J^2\cos^2\theta}{1+\Lambda_{(\mp)}}\right)^{\frac{1}{2}},
\label{eqn:potential})
\end{equation}
\begin{equation}
\Lambda_{(\pm)}=-\frac{mJr\sin\theta}{\mu^2r^2-{\cal S}^2}
\pm\left[\frac{\mu^2J^2r^2\sin^2\theta}{(\mu^2r^2-{\cal S}^2)^2}
-\frac{\left(J^2-\frac{2MJ^2}{r}\cos\theta-{\cal
S}^2f\right)}{\mu^2r^2-{\cal S}^2}\right]^{\frac{1}{2}}.
\label{eqn:potfunc})
\end{equation}
The particle with energy $E$ can move in the
region $E>V_{(\pm)}(r,\theta;J_z,{\cal S})$. 
$\pm$ in Eqs.(\ref{eqn:potential}) and (\ref{eqn:potfunc}) means the spin direction, up and down respectively. 
Up (down) means that the spin vector points in the same (opposite) direction on
the equatorial plane as the total angular momentum vector. 
We analyzed the motion of the particle with an up-spin\cite{footnote}. 

The ``effective potential'' is classified into four types. 
These are shown in Fig.2 and 3 in Paper I. 
It was shown that the only particle in type(B2) potential could be
chaotic. 

\section{Gravitational wave from a spinning particle}
\label{sec3}\setcounter{equation}{0}
In this section, the energy emission rates of the gravitational waves and
the wave forms from a spinning particle both in the regular and chaotic phases
are calculated by the quadrupole formula. 
Note that since the orbits considered in this study are very relativistic ones,
the quadrupole formula may provide only a qualitative picture.

\subsection{Motion of a spinning test particle}
In order to study the gravitational waves from a chaotic system, we analyze the three typical orbits, (a) an
``elliptic'' orbit restricted in the equatorial plane (quotation marks are used because this
orbit is not true elliptical but has the perihelion shift), (b) a quasi-periodic orbit leaving
from the equatorial plane and (c) a chaotic orbit in this paper.  These orbits are shown in Fig.1. 
Note that the coordinates $(\rho,z)$ used in
Fig.1 are defined as $\rho=r\sin\theta$ and
$z=r\cos\theta$. 
These particles have the same parameters, i.e. $E=0.94738162\mu$, $J_z=4\mu M$
and ${\cal S}=1\mu M$, which belong to the type(B2) potential, but different
initial conditions. 
The initial position are given as $r_0=7M$ for orbits (a) and (b) and $r_0=3.9M$ for orbit (c)
and initial spin direction for orbit (a) is perpendicular to the equatorial plane 
and determined in order that
$p_r$ vanishes initially those for orbit (b) and (c). 

The circular orbit with same parameters except the energy is used to normalize
the quantities calculated below, such as the amplitude and the energy spectrum
of the gravitational waves. 
The orbital radius for the circular orbit is $r=6.4843M$. 

\subsection{Energy emission rate}
According to the quadrupole formula, the averaged amplitude of the gravitational
wave for all directions is given as\cite{Landau}
\begin{equation}
A\propto\left(\frac{\mbox{d}^2Q_{ij}}{\mbox{d}t^2}\frac{\mbox{d}^2Q_{ij}}{\mbox{d}t^2}\right)^{1/2}.
\end{equation}
where $Q_{ij}$ is the reduced quadrupole moment of the system, 
given as
\begin{equation}
Q_{ij}\equiv
\mu\left(x_{i}x_{j}-\frac{1}{3}\delta_{ij}r^2\right),\;\;x_{i}=(x,y,z).
\end{equation}
The energy emission rate of the
gravitational waves is given as
\begin{equation}
\frac{\mbox{d}E}{\mbox{d}t}=\frac{1}{5}\left\langle\frac{\mbox{d}^3Q_{ij}}{\mbox{d}t^3}\frac{\mbox{d}^3Q_{ij}}{\mbox{d}t^3}\right\rangle,
\end{equation}
where $\langle\;\rangle$ denotes the average over several orbit.
For each case, the averaging time is several averaged orbital
periods of a spinning particle around a black hole. 
Because the angular momentum and the averaged orbital radius of the particles
are almost the same for each case, the averaging times are also the same for each case.
Therefore varying this averaging time does not change the following results. 
As we can see from this expression, the amplitude of gravitational waves gets large
when the mass distribution of the system changes violently.
Therefore, one may expect that the amplitude and the
energy emission rate of the gravitational waves from the chaotic
system would be larger than these from the quasi--periodic orbit.

So far, there were two works on the energy emission rate
and the amplitude of the gravitational waves for chaotic systems\cite{Kokubun},\cite{Cornish}.
In \cite{Kokubun}, the author calculated the energy emission rate from a
particle moving in the Henon-H\'eiles potential. 
In \cite{Cornish}, they analyzed the motion of a (spinless) particle in the 
Majumder--Papapetrou spacetime with two black holes. 
Comparing the amplitudes of the
gravitational waves both from the quasi--periodic orbit and from the chaotic one were calculated, they showed that the
amplitude and the energy emission rate were enhanced by a chaotic motion of a particle. 

In this paper, we discuss more generic motion of a particle, i.e. a
particle leaving from the equatorial plane, whereas the motion of a
particle in \cite{Kokubun} and \cite{Cornish} is restricted to single plane .

In Fig.2, we depict the time variations of the averaged amplitude and the energy emission rate for
the above  three orbits shown in Fig.1. 
Those amplitudes and emission rates are normalized by that for the circular orbit in the
same ``effective potential''. 
The maximum amplitudes for the orbits (a) and (c) are about 60\% times larger than that
of the circular orbit. 
The amplitudes for orbits (a) and (c) are nearly equal, but, as we can see, the amplitude for the chaotic orbit (c) is
slightly changing in time, whereas it is constant for the ``elliptic'' orbit (a). 

We also find that the energy emission rates for orbits (a) and (c) are about 10 times larger than
that for circular orbit. 
Notice that the energy emission rate and the amplitude for orbit (a) is larger
than that for orbit (c), i.e., the chaotic motion does not provide maximum enhancement of the energy. 
Rather, it may depend on how closely the particle approaches the black hole. 
In fact, the particle in chaotic orbit (c) has a
non-zero $z$ component of the momentum. 
Then, the particle cannot touch the edge of the ``effective potential'' in
general. 
This means that the minimum orbital radius of the orbit (c) is larger than that of the 
orbit (a). 
This is the reason that the energy emission rate and the amplitude for the ``elliptic'' orbit (a)
is slightly larger than that for the chaotic orbit (c).
The orbital radius of orbit (b) does not change largely and the minimum radius is largest
comparing with the other cases. 
This is the reason why the amplitude and the energy emission rate is smaller than those for orbit
(a) and (c). 

\subsection{The wave forms and the energy spectrum}
As we have seen in the previous subsection, the amplitude and the emission rate are not a good variable to observe a chaotic
motion. 
Then, we will analyze the wave forms or its energy spectrum, which may show some qualitative difference for a chaotic motion. 
In the cases (b) and (c), the orbital plane of the particle precesses. 
The time variation of the polar angle of
the instantaneous orbital plane is shown in Fig.3 for the orbits (b) and (c). 
The polar angle $\Theta$ of the instantaneous orbital plane is defined by
\begin{equation}
\cos\Theta\equiv \sqrt{\frac{L_x^2+L_y^2}{L_z^2}},
\end{equation}
\begin{equation}
L_{\mu}\equiv f_{\mu\nu}p^{\nu},
\end{equation}
where $f_{\mu\nu}$ is the Killing--Yano tensor of the background
spacetime and $L_i$ is interpreted as the orbital angular momentum of the
particle\cite{Rudiger}.

We can see the clear difference between them. 
This different behavior of the orbital plane will change those wave forms detected by
the observer.

The gravitational wave has two modes $h_{+}$ and $h_{\times}$ in
TT-gauge\cite{Landau}, which are given as
\begin{eqnarray}
h_{+}&=&\frac{h_{\theta\theta}}{r^2}=\left[\left(h^{\mbox{\tiny{Q}}}_{xx}-h^{\mbox{\tiny{Q}}}_{yy}\right)\frac{(\cos^2\theta+1)}{4}\cos2\phi-\frac{\left(h^{\mbox{\tiny{Q}}}_
{xx}+h^{\mbox{\tiny{Q}}}_{yy}-2h^{\mbox{\tiny{Q}}}_{zz}\right)}{4}\sin^2\theta\right.\nonumber\\
&&\left.+h^{\mbox{\tiny{Q}}}_{xy}\left(\frac{\cos^2\theta+1}{2}\right)\sin2\phi-h^{\mbox{\tiny{Q}}}_{xz}\sin\theta\cos\theta\cos\phi-h^{\mbox{\tiny{Q}}}_{yz}\sin\theta\cos\theta\sin\phi\right],\nonumber\\
h_{\times}&=&\frac{h_{\theta\phi}}{r^2\sin^2\theta}=-\frac{(h^{\mbox{\tiny{Q}}}_{xx
}-h^{\mbox{\tiny{Q}}}_{yy})}{2}\cos\theta\sin
2\phi+h^{\mbox{\tiny{Q}}}_{xy}\cos\theta\cos 2\phi\nonumber\\
&&+h^{\mbox{\tiny{Q}}}_{xz}\sin\theta\sin\phi-h^{\mbox{\tiny{Q}}}_{yz}\sin\theta\cos\phi,
\label{eqn:quadrupolewave}
\end{eqnarray}
where $h^{\mbox{\tiny{Q}}}_{ij}$ is defined as
\begin{equation}
h^{\mbox{\tiny{Q}}}_{ij}=\frac{2}{r}\frac{\mbox{d}^2Q_{ij}}{\mbox{d}t^2}.
\end{equation}
 and $(r,\theta,\phi)$ indicates the position of the observer. 
The waveforms of $h_{+}$ and $h_{\times}$ are shown in Fig.4, in which the
observer is assumed to be on the equatorial plane.
The top figure shows the waveform from the orbit (a), middle figures show those from the orbit (b)
and bottom figures for the orbit (c). 
Note that $h_{\times}$ for the orbit (a) vanishes because the particle moves on the equatorial plane and the observer is also on the
equatorial plane. 
For the orbits (a) and (c), the orbital radius changes from $r\simeq3.9M$ to $r\simeq 13M$ (Fig.1). 
Then the small and large peaks of the gravitational waves appear alternately.  
The difference in $h_{+}$ for the orbits (a) and (c) is not so clear. 
However, for the orbit (c), the irregular variations of wave forms can be found, especially in the variation
of the wave form for small peaks. 
The apparent difference in $h_{\times}$ for the orbits (b) and (c) comes from that in the behavior of the orbital plane
(see Fig.3). 
This might be a sign of a chaotic motion, although we need more detailed analysis. 

The energy spectrum of the gravitational wave is one of the most important observables in gravitational
astronomy. 
The energy spectrum of the gravitational wave is calculated as
\begin{equation}
\frac{\mbox{d}E}{\mbox{d}\omega}\propto
(|\hat{A}_{+}|^2+|\hat{A}_{\times}|^2),
\end{equation}
where $\hat{A}$ denotes the Fourier coefficients of each wave mode. 
Note that the time interval for integration to obtain the Fourier coefficient numerically is much
longer than the inverse of the Lyapunov exponent for chaotic case and for the other cases, it is
taken in order that the number of rotation is approximately equal to it for chaotic case. 
  In Fig.5, the energy spectra
for each case are shown. These are normalized by the peak value in the circular orbit shown in
Fig.5(i).  Note that for the circular orbit, the spectrum has
only one sharp peak at $\omega=0.1148\;[c^3/GM]$ which corresponds to twice of the orbital frequency. 

The spectra of the orbits (a) and (c) show
many peaks which never appear in the circular orbit.  The intervals between each peak are almost
constant.  They are approximately $\Delta\omega\simeq 0.025\;[c^3/GM]$. 
This frequency corresponds to it for the change of the orbital radius and is approximately a half
of orbital frequency, i.e., the perihelion shift of the particle is $2\pi$. 
For orbit (b), we can see additional peaks in the spectrum. 
But the region where these peaks appear is narrow and the entire shape is similar to the circular
case because the variation of the orbital radius is small and the average radius is almost equal
to that of circular orbit.
 For the orbit (c), we
find many small spikes in entire region.  In the low frequency region, the width of the spikes is
broadened compared with other cases. 
The signal contains all frequency components besides in addition to the peak
frequencies appear in the quasi-periodical case. 
 This character in the Fourier spectra
means that the signal is not (quasi-)periodical, which is known as the peculiar character of the
chaotic system\cite{Lichtenberg}.  To see this feature more clearly, we show the energy spectrum
of the gravitational waves from more chaotic case in Fig.6.
In this case, the magnitude of spin is given as ${\cal S}=1.2\mu M$. 
Note that this value makes the motion of particle more chaotically but unphysical (see, Paper I). 
For orbit (c), we can see the character of chaos mentioned above more clearly.
These features are thought to be caused mainly by the irregular time variation of the orbital
radius shown in Fig.7 rather than the precession of the orbital plane.

\section{Summary and Discussions}
\label{sec4}\setcounter{equation}{0}
In this paper, we investigate a motion of a spinning test particle
around a Schwarzschild black hole and estimate the gravitational
waves emitted from such a system by quadrupole formula. 
For typical four representative orbits, a circular orbit, an 
``elliptic'' orbit on the equatorial plane, a quasi-periodic orbit, and
chaotic orbit, the average amplitude of gravitational waves, energy emission rates, wave forms and the energy
spectra of the gravitational waves are calculated.

It has been expected that the energy emission rate will be enhanced by the
chaotic motion because the energy emission rate is, closely related to a time variation of the mass
distribution of the system.
The numerical results show that the average amplitude and the energy emission rate of the gravitational waves for the chaotic orbit 
are by 60\% and about 10 times larger than those for the circular orbit, respectively. 
But, those are slightly smaller than those for the regular ``elliptic'' orbit. 
For the ``elliptic'' orbit, the particle reaches the inner edge of the ``effective potential'', 
where the energy emission rate becomes largest. 
On the other hand, for the chaotic orbit, the particle has a momentum component perpendicular to
the equatorial plane, and then the particle in general does not touch the edge of the ``effective
potential''.
The particle is reflected near the inner edge of the ``effective
potential''.
Thus, the energy emission rate gets slightly smaller.
However, the variation of the amplitude and the energy emission rate for
the chaotic orbit is clearly different from the regular
one.

The quadrupole formula is also used to obtain the wave forms. 
It has the same angle--dependence as the spherical harmonics of $l=2$.
Because the orbital plane precesses for orbits(b) and (c) due to the spin as shown in Fig.2, the wave
forms observed by a fixed observer are expected to be modified through precession.
The behavior of the orbital plane for the chaotic case is obviously different from that for the regular
case and the wave form for the  chaotic case especially $h_{\times}$ is also different from that for the regular case (Fig.3).

The energy spectra, which are important observables in gravitational wave astronomy, are also
calculated for each case.
For the chaotic orbit, the spectra becomes noisy because of the appearance of small spikes in entire
region and the large spikes in low frequency region are broadened compared with those in the regular case.

Finally, we comment on the possibility of detecting chaos occurring in a real binary system and the
effects on  gravitational wave astronomy.
The time scale for which the chaos becomes conspicuous is indicated by
$\lambda^{-1}$, where
$\lambda$ is the Lyapunov exponent of the system.
For our model, $\lambda$ is about several times the average orbital
period (Paper I).
The time scale for which the system changes its state is indicated by
$E/\mbox{d}E/dt$, which is about 10 times the orbital period.
This means that the system has enough time to show a chaotic behavior, if the chaos once occur in the system.
In the real astrophysical system, there are other components which may induce a chaotic motion besides a spin.
For example, in \cite{Letelier}, it is shown that the higher multipole
moments of a black hole induce chaos in a motion
of a particle.
In \cite{Cornish}, the motion of the particle around a extreme
Reissner-Nordostr\"om black hole perturbed by a tiny gravitating
source was analyzed and it was shown that chaos can occurs.
Thus, chaos may occur easily in a real astrophysical system.
Then it is natural that the binary system is always perturbed by other
gravitating sources and it might show some chaotic behavior to be observed.
The correlation function method used in \cite{Lehto}--\cite{Kollath} may be useful to judge
whether the signal is generated by the stochastic system (noise) or the chaotic system with finite
dimension. 

In addition, it was shown
that a spin interaction would change the radius of the ISCO and at the final moment of the
inspiral phase, the orbital plane may precess violently, as in Paper II. The parameters which
induce this instability belong to the type(B2) potential for which chaos can occur.
If such peculiar signatures are found in the detected
signal of the gravitational waves, we may be able to extract some additional information about the
source, for example a lower bound of a spin.

In future works, we will analyze other relativistic effects, for
example quadrupole moments of a particle and of a central object, on the orbit of the particle and the emitted gravitational waves.
We should also study in detail how to extract new information
about a chaotic system, if it exists, from the detected signal.
The quadrupole formula used in this paper should, of course, not be used
for calculating the gravitational waves emitted from relativistic system.
It may give us only qualitative results. 
Therefore, we have to use a relativistic method, such as a black hole perturbation
method, to calculate more accurate gravitational waves and discuss the effect of chaos on the
observation of the gravitational waves. 
It may require to include a back reaction effect\cite{Mino2}.

\acknowledgements{
We would like to thank Paul Haines for his critical reading  of
our paper. This work was supported partially by the Grant-in-Aid
for Scientific Research Fund of the Ministry of Education,
Science and Culture (Specially Promoted Research No. 08102010),
by a JSPS Grant-in-Aid (No. 095790),
and by the Waseda University Grant
for Special Research Projects.}



\newpage
\begin{flushleft}
{ Figure Captions (ps-figures are found in http://www.phys.waseda.ac.jp/gravity/\~{}shingo)} 
\end{flushleft}

\vskip 0.1cm
   \noindent
\parbox[t]{2cm}{ FIG. 1:\\~}\ \
\parbox[t]{14cm}
{The typical orbits treated in this paper in $r$-$z$ plane:(a) the ``elliptic''
orbit restricted on the equatorial
plane, (b) the quasi--periodical orbit and (c) the chaotic orbit.
They have same parameters $E=0.94738162\mu$, $J=4\mu M$ and ${\cal S}=1\mu
M$ but different initial condition, which
are (a) $r_0=7M$ and $\eta=\pi/2$ (b) $r_0=7M$ and (c) $r_0=3.9M$,
respectively.
$\eta$ for orbit(b) and (c) is determined from $p_r=0$.}\\[1em]
\noindent
\parbox[t]{2cm}{ FIG. 2:\\~}\ \
\parbox[t]{14cm}
{(i) The average amplitude of the gravitational waves for all direction for
orbit (a) (dotted line), orbit(b) (dashed
line) and orbit (c) (solid line). 
(ii) The energy emission rate of the gravitational waves.  
These are normalized by the value for
the circular orbit in the same ``effective potential'', which is constant. 
The amplitude and emission rate for
orbit (a) is slightly larger than those for the chaotic case.}\\[1em]
\noindent
\parbox[t]{2cm}{ FIG. 3:\\~}\ \
\parbox[t]{14cm}
{The precession of the orbital plane for (i) orbit(b) (quasi--periodical
case) and (ii) orbit(c) (chaotic case).}\\[1em]
\noindent
\parbox[t]{2cm}{ FIG. 4:\\~}\ \
\parbox[t]{14cm}
{The gravitational wave forms of the plus mode and the cross mode for the
observer on the equatorial plane.
The top panel shows the gravitational wave for orbit (a). The cross mode vanishes. 
The middle panels show the the gravitational waves for orbit (b). 
The bottom ones show these for orbit(c).}\\[1em]
\noindent
\parbox[t]{2cm}{ FIG. 5:\\~}\ \
\parbox[t]{14cm} {The energy spectra of the gravitational wave for each case.
(i) The energy spectra for the circular orbit. 
The frequency of the peak is $\omega=0.1148\;[c^3/GM]$. 
(ii)-(iv) The energy spectrum for each orbit. 
The values in these cases are normalized by the peak value for the circular orbit. 
The small boxes attached to each figures show the enlargement of the low frequency region.
For orbit (c) (chaotic), the spectra is noisy and each spikes are broadened compared with the other
cases.  
}\\[1em]
\noindent
\parbox[t]{2cm}{ FIG. 6:\\~}\ \
\parbox[t]{14cm} {The energy spectra of the gravitational wave from a spinning particle with
${\cal S}=1.2\mu M$, $J=4 \mu M$ and $E=0.93545565\mu$. 
The types of orbit are same as those in ${\cal S}=1 \mu M$ case. 
In (iv) (chaotic case), we can see that the small spikes exist and the peaks appear in other cases
are broadened. }\\[1em]
\noindent
\parbox[t]{2cm}{ FIG. 7:\\~}\ \
\parbox[t]{14cm} {Time variations of orbital radius for each cases. 
The principle frequency is about half of the orbital frequency.  
}\\[1em]
\noindent

\begin{thebibliography}{99}
\bibitem{footnote} the definition of the sign of the spin direction is different from that in Paper I. 
The effective potential $V_{(+)}$ is same as the $V_{(-)}$ in Paper I.
\end{thebibliography}
\end{document}